# Fast time-domain current measurement for quantum dot charge sensing using a homemade cryogenic transimpedance amplifier

Heorhii Bohuslavskyi[1,a)], Masayuki Hashisaka[1,b)], Takase Shimizu[1,2], Takafumi Akiho[1], Koji Muraki[1], and Norio Kumada[1]

[1]NTT Basic Research Laboratories, NTT Corporation, 3-1 Morinosato-Wakamiya, Atsugi, Kanagawa 243-0198, Japan

[2]Institute for Solid State Physics, The University of Tokyo, 5-1-5 Kashiwanoha, Kashiwa, Chiba 277-8581, Japan

[a)]Current address: VTT Technical Research Centre of Finland, Tietotie 3, 02150 Espoo, Finland

[b)]Author to whom correspondence should be addressed: masayuki.hashisaka.wf@hco.ntt.co.jp

We developed a high-speed and low-noise time-domain current measurement scheme using a homemade GaAs high-electron-mobility-transistor-based cryogenic transimpedance amplifier (TIA). The scheme is versatile for broad cryogenic current measurements, including semiconductor spin-qubit readout, owing to the TIA's having low input impedance comparable to that of commercial room-temperature TIAs. The TIA has a broad frequency bandwidth and a low noise floor, with a trade-off between them governed by the feedback resistance $R_{FB}$. A lower $R_{FB}$ of 50 k$\Omega$ enables high-speed current measurement with a $-3$dB cutoff frequency $f_{-3dB} = 28$ MHz and noise-floor $NF = 8.5 \times 10^{-27}$ A$^2$/Hz, while a larger $R_{FB}$ of 400 k$\Omega$ provides low-noise measurement with $NF = 1.0 \times 10^{-27}$ A$^2$/Hz and $f_{-3dB} = 4.5$ MHz. Time-domain measurement of a 2-nA peak-to-peak square wave, which mimics the output of the standard spin-qubit readout technique via charge sensing, demonstrates a signal-to-noise ratio (SNR) of 12.7, with the time resolution of 48 ns, for $R_{FB} = 200$ k$\Omega$, which compares favorably with the best-reported values for the radio-frequency (RF) reflectometry technique. The time resolution can be further improved at the cost of the SNR (or vice versa) by using an even smaller (larger) $R_{FB}$, with a further reduction in the noise figure possible by limiting the frequency band with a low-pass filter. Our scheme is best suited for readout electronics for cryogenic sensors that require a high time resolution and current sensitivity and thus provides a solution for various fundamental research and industrial applications.

There is a growing demand for fast and precise readout electronics for cryogenic sensor devices over a broad range of scientific research and future quantum technologies [1-7]. Spin-qubit readout is one of the most demanding applications for such extreme electronics. While RF reflectometry technique has been explored for high-speed measurements on large-scale qubit arrays [1,2], applying these techniques to new materials and devices with unknown properties is challenging since it requires a highly elaborate design of the measurement setups. On the other hand, a broadband cryogenic amplifier enables direct current (DC) measurement with a high time resolution and current sensitivity, another promising method for the spin-qubit readout [8-12]. While the finite power consumption of a cryogenic amplifier limits the number of measurement lines in a cryostat, preventing its application for future large-scale qubit arrays, the technique provides a simple and versatile measurement method for small-scale systems in high demand for basic research and various early-stage technology developments. Indeed, several recent spin-qubit experiments have used the cryogenic-amplification technique [11-12], and the development of faster and more precise setups is continuously advancing [9-12]. Despite intensive studies on high-speed time-domain measurements, however, the measurement bandwidth obtained using a conventional voltage amplification is limited to a few megahertz. Thus, demonstrating an alternative cryogenic-amplification approach for a better time resolution is a timely challenge in applied physics.

A cryogenic common-source circuit composed of a high-electron-mobility transistor (HEMT) has often been used as readout electronics for spin-qubit charge sensing [9-12]. In this conventional setup, current output from a charge sensor flows through a resistor located at the lowest temperature to generate a voltage, which is then amplified by the common-source circuit before it has been transferred to the room-temperature equipment. Although this technique has enabled precise charge sensing with a time resolution of about a few hundred nanoseconds [11-12], it has a fundamental limitation caused by the resistance-capacitance (RC) damping before the cryogenic amplification to achieve even higher measurement speeds. The parasitic capacitance of a coaxial cable connecting the device-under-test (DUT) and the voltage amplifier forms an RC low-pass filter (LPF) with the resistor used for the cryogenic current-voltage conversion [3]. Although previous studies have examined placing the voltage amplifier close to the DUT to reduce the cable length and hence the parasitic capacitance, the measurement frequency band is still limited to below a few megahertz [9-12]. In addition, such implementations have several practical drawbacks, such as an increase in the temperature of the DUT and reduced versatility of the measurement setup.

This letter describes a time-domain measurement scheme using a homemade cryogenic transimpedance amplifier (TIA). The TIA simultaneously provides broadband current-to-voltage conversion and low-noise cryogenic signal amplification, which makes the scheme suitable for high-speed and precise current readout. Due to the low input impedance of the TIA, the RC damping at the input does not influence the time-domain response, unlike in the common-source circuit. Instead, what limits the frequency bandwidth of the current readout is the −3dB cutoff frequency $f_{-3dB} \approx (2\pi R_{FB} C_{FB})^{-1}$ that is mainly determined by the resistance $R_{FB}$ and capacitance $C_{FB}$ in the feedback loop. The noise floor of the measurement is given as the input-referred noise of the TIA, which is dominated by the thermal noise $4k_B T R_{FB}$ ($k_B$ is Boltzmann's constant; $T$ is temperature) at megahertz frequencies. As we show below, by appropriately choosing $R_{FB}$, the noise of the TIA can be reduced to where it is low enough for charge sensing, even when it operates at 4.2 K. There is a trade-off between the response time and noise floor governed by $R_{FB}$. Therefore, the $R_{FB}$ value should be chosen to achieve the best performance for each purpose. The power consumption of our TIA is about 1.5 mW, which is low enough to use it at the 4-K or still stage of a commercial dilution refrigerator. We demonstrate the TIA's performance by measuring 2-nA peak-to-peak square-wave signals, which mimic the output of a standard charge sensing for qubit readout. The signal-to-noise ratio (SNR) for $R_{FB}$ = 200 kΩ is 12.7, with a time resolution of 48 ns, which is comparable to the state-of-the-art RF reflectometry technique readout [13-15]. Thus, our TIA has excellent potential for industrial and fundamental research applications requiring broadband and low-noise cryogenic amplification.

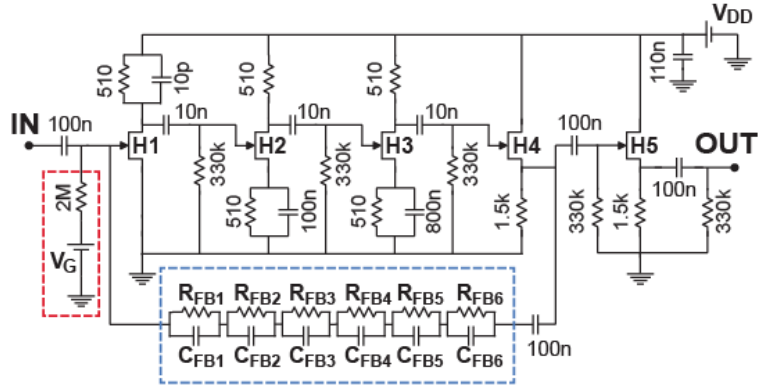

**Figure 1.** TIA electronic circuit schematic. In addition to the HEMTs, the printed circuit board (PCB) is populated with standard thin-film SMD capacitors and resistors. The TIA has two supply voltage lines, $V_G$ and $V_{DD}$. Red and blue dashed squares highlight the main differences from our previous TIA [19].

Figure 1 shows the electric circuit diagram of the TIA. The amplifier has two control voltages, $V_{DD}$ and $V_G$, capacitively coupled input and output, and a cold ground line. The circuit consists of three common-source amplifying stages (H1, H2, and H3) followed by source-followers before (H4) and after (H5) the feedback line, which gives the low output impedance $Z_{out} \approx 30\ \Omega$. Here, H1-H5 labels the five homemade HEMT chips, whose channel width $W = 3$ mm and gate length $L_g = 4$ µm [16]. While the circuit design follows that of the TIA developed for shot-noise measurement on a mesoscopic sample [17-19], we made two significant improvements to make it faster and more sensitive, aiming at time-domain measurements. First, unlike H2 and H3 in the second and third amplifying stages, which are self-biased, we employ an additional voltage source $V_G$, exclusively to bias H1 in the first stage (highlighted by the red dashed square in Fig. 1) to optimize its operation. As the first stage governs the input-referred noise of the TIA, the fine-tuning of H1's operation point is most effective in reducing the noise floor. Second, instead of having one resistor with a feedback resistance $R_{FB}$, we use a chain of six resistors $R_{FB1} - R_{FB6}$ with the total resistance $R_{FB} = \Sigma_i R_{FBi}$ in the feedback loop (highlighted by the blue dashed square). Splitting a single feedback resistor reduces the effective feedback capacitance due to the parasitic capacitance of the surface-mounted-device (SMD) resistor. The reduced feedback capacitance significantly increases $f_{-3dB}$, allowing a faster measurement, while it remains sufficiently large for the phase compensation to stabilize the TIA operation. Our TIA typically operates at $V_{DD} \approx 0.7$ V and $V_G \approx -0.31$ V, where the power consumption is about 1.5 mW. The power consumption can be reduced by decreasing $V_{DD}$ at the cost of noise performance [16]. The input impedance $Z_{in}$ is typically about 100 $\Omega$ with these voltage biases, providing stable operation against changes in DUT resistance $R_{DUT}$ and the parasitic capacitance of a coaxial cable, hence allowing versatile use of the TIA for various applications (For more details, see the Supplementary Material.)

We evaluated the TIA's transimpedance and noise properties with four different $R_{FB}$ values (50, 100, 200, and 400 k$\Omega$) in the frequency domain by connecting a test resistor $R_{DUT} = 100$ k$\Omega$ with an adaptor. In this case, the input capacitance $C_{IN}$ between the DUT and TIA is about a few picofarads. In this characterization, the TIA cooled down to 4.2 K converts a sinusoidal alternating current (AC) of $I_{in} = 10$ nA$_{rms}$ to an output voltage $V_{out} = Z_{trans} I_{in}$ with its specific transimpedance $Z_{trans}$.

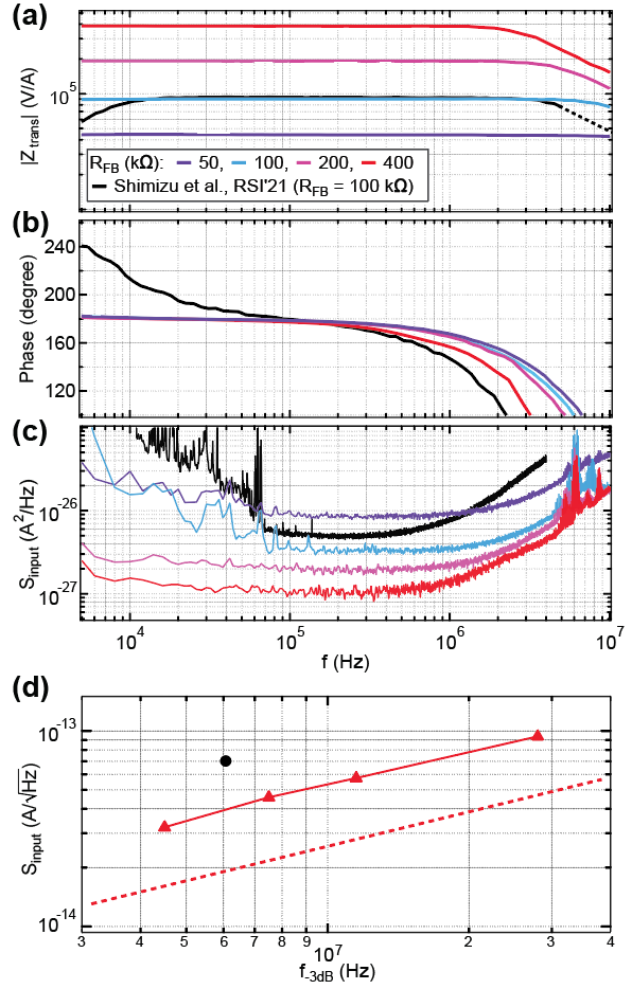

**Figure 2.** (a) Absolute value of transimpedance $|Z_{trans}|$, (b) phase, and (c) $S_{input}$ of TIAs plotted as a function of frequency measured at 4.2 K for a DUT of $R_{DUT} = 100$ kΩ. The data of our previous TIA are shown in black [19]. (d) Summary of minimum $S_{input}$ (vertical axis) and $f_{-3dB}$ (horizontal axis) at 4.2 K (red triangles), plotted with those of our previous TIA (black circle [19]). The dashed line indicates the expected performance of our TIA at 1 K.

A room-temperature commercial voltage amplifier (NF corporation SA-430F5) amplifies the output voltage, and then a digitizing oscilloscope (National Instruments PXI-5122) measures the signal for the following FFT processes. The Bode plots in Figs 2(a) and (b) show the evaluated frequency dependence of $|Z_{trans}|$ and the phase shift, respectively. The input-referred current noise $S_{input}$ spectra, namely the noise floor of the setup with the TIA, are presented in Fig. 2(c). The results of the previous study with $R_{FB} = 100$ kΩ [19] are added for comparison. The present TIAs show flat $|Z_{trans}|$ over the broad frequency bands required for accurate time-domain measurements. The bandwidth increases with decreasing $R_{FB}$; from the data in Fig. 2(a), $f_{-3dB}$ is estimated to be 11.5, 7.5, and 4.5 MHz for $R_{FB} = 100$, 200, and 400 kΩ, respectively. For $R_{FB} = 50$ kΩ, at which $|Z_{trans}|$ remains flat up to the highest frequency, we used a time-domain measurement described later to find $f_{-3dB} = 28$ MHz as $0.35/\tau_{rise}$, where $\tau_{rise}$ is the 10–90% rise time. Note that $f_{-3dB} = 11.5$ MHz obtained for $R_{FB} = 100$ kΩ is almost doubled compared with that reported previously for the same $R_{FB}$ [19]. As shown in Fig. 2(b), the phase stability is also improved for all four TIAs, staying close to the 180° ideal for negative feedback over a wide frequency range.

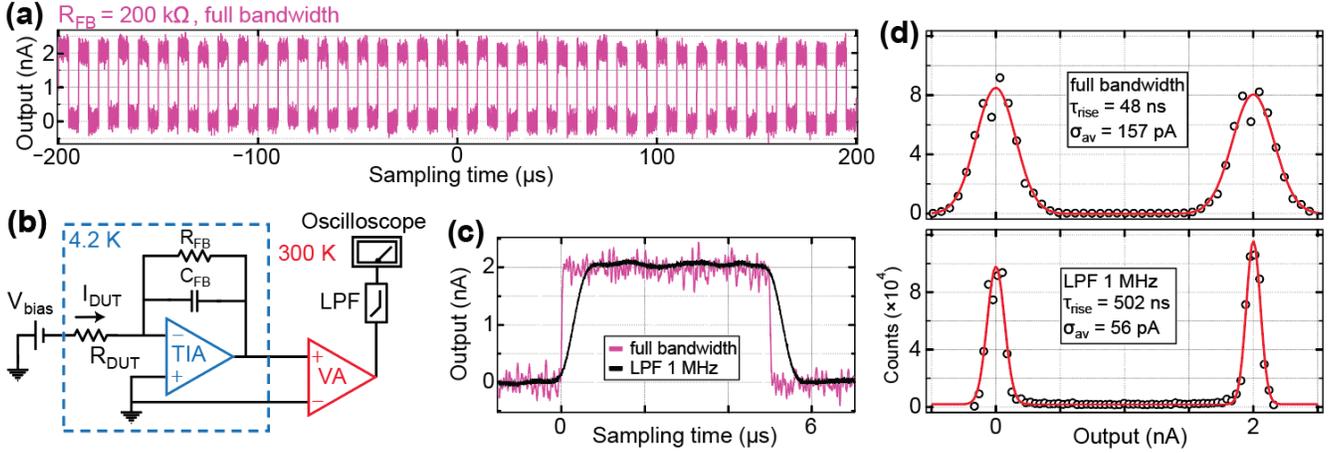

**Figure 3.** (a) Single-shot trace for 2 nA peak-to-peak square-wave signal measured with $R_{FB}$ = 200 kΩ TIA. (b) Measurement setup. The output voltage from TIA is amplified by room-temperature amplifier (VA) and measured by oscilloscope with or without an external LPF. (c) Single-shot traces recorded at full bandwidth (pink) and with 1-MHz LPF (black). (d) Histogram analyses for single-shot data measured at full bandwidth (upper) and with 1-MHz LPF (lower). The data are fitted with a double-Gaussian function.

Besides the broader TIA bandwidth, we observe in Fig. 2(c) a white low noise floor spanning from 100 kHz up to approximately 1 MHz. For these plots, we subtracted the Johnson–Nyquist noise contribution $4k_BT/R_{DUT}$ = 2.38×10$^{-27}$ A$^2$/Hz of $R_{DUT}$ = 100 kΩ at 4.2 K. The present TIA with $R_{FB}$ = 100 kΩ shows an improved noise floor $NF$ ≈ 3.5×10$^{-27}$ A/Hz$^{1/2}$, a 30% reduction compared with the previous report [19]. The noise floor between 100 kHz and 1 MHz, decreasing almost linearly with increasing $R_{FB}$, is close to the thermal noise $4k_BT/R_{FB}$ of the feedback resistance, indicating that the thermal noise dominates the noise floor over the HEMT's input voltage and current noise. The characteristic frequency $f_C$, above which the noise amplitude increases with frequency, due to the increased noise gain, is about one megahertz. The $f_C$ value is almost doubled compared with the previous TIA [19], indicating a lower-noise performance at high frequencies. Figure 2(d) summarizes $f_{-3dB}$ and minimal $S_{input}$ of our TIAs. The time resolution and noise floor trade-offs associated with the $R_{FB}$ value. The $R_{FB}$ = 50 kΩ TIA with $f_{-3dB}$ = 28 MHz and $NF$ = 8.5 × 10$^{-27}$ A$^2$/Hz is suitable for high-speed measurement, while the 400 kΩ one with $f_{-3dB}$ = 4.5 MHz and $NF$ = 1.0 × 10$^{-27}$ A$^2$/Hz is suitable for high-precision measurement. Since $S_{input}$ of our TIAs is dominated by the thermal noise of $R_{FB}$ at 4.2 K, we can expect even better performance when they are operated at a lower temperature, e.g., at the still stage (1K), as indicated by the dashed line.

Now, we turn to time-domain measurements with the $R_{FB}$ = 200 kΩ TIA using a digital oscilloscope (Tektronix DPO3032). Figure 3(a) shows a single-shot trace for a continuous 2-nA peak-to-peak square wave through the 100-kΩ-$R_{DUT}$ sample, acquired with the full bandwidth of the TIA, clearly indicating the great potential of the TIA for time-domain measurement. We evaluated the time resolution and noise properties for this measurement setup, schematically shown in Fig. 3(b), with and without an auxiliary LPF installed at the oscilloscope's input. Figure 3(c) compares the full-bandwidth data in Fig. 3(a) and the data recorded with a 1-MHz LPF for a single square-wave pulse. While the LPF increases the 10–90% rise time $\tau_{rise}$ from 48 to 502 ns, it significantly reduces the background fluctuations to improve the current sensitivity. We evaluated the measurement noise by averaging the standard deviations $\sigma_0$ and $\sigma_1$ of the two current levels as $\sigma_{av}$ =

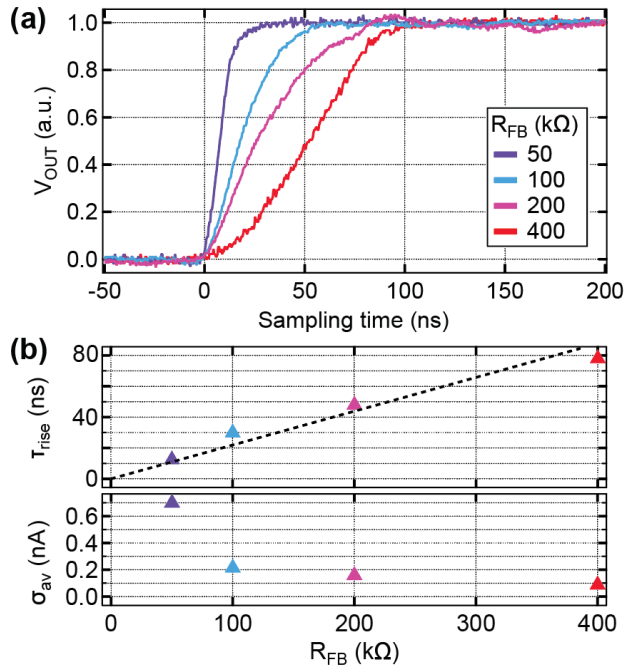

**Figure 4.** (a) Time traces acquired with TIAs. (b) Summaries of $\tau_{\text{rise}}$ (upper) and $\sigma_{\text{av}}$ (lower) of different-$R_{\text{FB}}$ TIAs at full bandwidth. The black dashed line is a linear fit for $\tau_{\text{rise}}$.

$\sqrt{(\sigma_0^2 + \sigma_1^2)/2}$, where $\sigma_0$ and $\sigma_1$ are estimated from the double-Gaussian fit for the histogram data, as shown in Fig. 3(d), and found that the LPF reduces $\sigma_{\text{av}}$ from 157 pA to 56 pA. Note that the thermal noise of the $R_{\text{DUT}} = 100$ k$\Omega$ resistor at 4.2 K contributes to $\sigma_{\text{av}}$; hence, the noise floor of the present setup will be even smaller in an actual charge-sensing experiment with the DUT at the mixing-chamber temperature. If we define the SNR of the present measurement as SNR = $\Delta/\sigma_{\text{av}}$, where $\Delta$ is the distance between the centers of the two Gaussian peaks in the histogram, we obtain SNR = 12.7 for the full-bandwidth measurement and SNR = 35.7 with the 1-MHz LPF.

We performed similar time-domain measurements and analyses for four TIAs with different $R_{\text{FB}}$ values. The time traces of square-pulse signals with normalized peak-to-peak amplitude in the full-bandwidth measurements are compared in Fig. 4(a). The TIAs of $R_{\text{FB}}$ = 400, 200, 100, and 50 k$\Omega$ show $\tau_{\text{rise}}$ of 78, 48, 30, and 12.5 ns, respectively, as summarized in the upper panel of Fig. 4(b). The almost linear $R_{\text{FB}}$ dependence of $\tau_{\text{rise}}$ reflects the TIA's bandwidth described as $f_{-3\text{dB}} \approx (2\pi R_{\text{FB}} C_{\text{FB}})^{-1}$. From the fit, we extracted the effective $C_{\text{FB}}$ of approximately 10–15 fF. On the other hand, $\sigma_{\text{av}}$ decreases with increasing $R_{\text{FB}}$, as shown in the lower panel, demonstrating the trade-off between the measurement speed and the current sensitivity concerning the $R_{\text{FB}}$ value. As discussed above, $\sigma_{\text{av}}$ can be suppressed by adding an auxiliary LPF at the cost of time resolution (see Supplementary Fig. S5).

While the data presented above were obtained with $C_{\text{IN}}$ of a few picofarads, we also tested our TIAs having $C_{\text{IN}}$ = 30 pF (corresponds to the parasitic capacitance of an approximately 50-cm-long coaxial cable) and found that the increased $C_{\text{IN}}$ affects the performance only slightly (see Supplementary Material). Therefore, while lowering $C_{\text{IN}}$ is important for maximum performance, the degradation by a finite $C_{\text{IN}}$ in standard wiring will not be a problem for most applications. We note that the impact of $C_{\text{IN}}$ on the bandwidth of charge sensing will be minimal for hot silicon spin qubits operated above 1 K [20-22], as the distance from the charge sensor and $C_{IN}$ can be small. Since the bandwidth and noise floor of RF reflectometry significantly degrade at 1 K or above [23], our amplifiers would offer unprecedented performance for hot spin qubits.

It is worth comparing the performance of our TIA-based readout electronics with that reported for RF reflectometry techniques. Our setup's bandwidth is much broader than the state-of-the-art reflectometry readout technique, which ranges from a few hundred kilohertz to ten megahertz at an SNR ≥ 1 [13-15]. Concerning the current sensitivity, the fastest reported reflectometry setup ($f_{-3dB}$ = 10 MHz set by an auxiliary LPF) has an SNR of 12.7 at 60 mK using the input power corresponding to excitation of approximately 2 nA$_{rms}$ [13]. It is not straightforward to directly compare the SNR of lock-in reflectometry homodyne detection with our observations. However, looking at our obtained values, e.g., the SNR = 12.7 for the 2-nA peak-to-peak signal with $f_{-3dB}$ = 7.5 MHz for the $R_{FB}$ = 200 kΩ, and reaching an SNR = 3 for $f_{-3dB}$ = 28 MHz ($R_{FB}$ = 50 kΩ), we expect the current sensitivity of our setup to be comparable to the best reflectometry technique readout. In addition, we also note that our TIA with $R_{FB}$ = 400 kΩ shows a similar $\sigma_{av}$ with a tenfold broader measurement bandwidth as compared with the recent common-source-amplifier-based readout circuit [11].

Finally, while the focus of this study is time-domain measurements, the advantage of our TIA for shot noise measurements is worth mentioning. When $R_{FB}$ = 400 kΩ is used, the average input-referred noise $S_{input-av}$ in the measurement bandwidth of $\Delta f$ = 900 kHz (from 0.1 to 1 MHz) is about $1.17 \times 10^{-27}$ A$^2$/Hz. With the data integration time of $\tau_{int}$ = 10 s, we expect the resolution of an autocorrelation current-noise measurement to reach $S_{input-av}/\sqrt{(\Delta f \times \tau_{int})} \approx 3.9 \times 10^{-31}$ A$^2$/Hz, which is better than that achieved in the previous TIA-based setup [19] and that in a conventional setup using a cryogenic voltage amplifier and an RLC resonator [16,24].

In summary, this letter presented a time-domain current measurement scheme using a cryogenic TIA based on homemade HEMTs. The TIA achieves high performance by optimizing the operating point of the first amplifying stage and suppressing the influence of the parasitic capacitance of the feedback resistance. The time-domain measurement demonstrated a record high time resolution and a high current sensitivity, which compares favorably with the results reported for the state-of-the-art RF reflectometry technique for spin-qubit charge sensing. While all the results were taken at 4.2 K, the TIA's low power consumption allows installing it on the still plate ($T \approx$ 1 K) in a dilution refrigerator to improve the current sensitivity further. The TIA is capable of various cryogenic current measurements in time and frequency domains, e.g., spin-qubit readout and shot-noise measurements.

**Supplementary Material**

See supplementary material for additional information about the HEMTs, the TIA footprint and design, the feedback resistance chains, the stability and reproducibility of TIAs, and the effects of stray input capacitance and low-pass-filtering on the TIA performance.

**Acknowledgements**

The authors thank H. Murofushi and M. Imai for technical support. This study was supported by the Japan Society for the Promotion of Science KAKENHI (Grant No. JP19H05603, JP21H01022, and JP22H00112).

# Supplementary Material for
# "Fast time-domain current measurement for quantum dot charge sensing using a homemade cryogenic transimpedance amplifier"

Heorhii Bohuslavskyi[1,a)], Masayuki Hashisaka[1,b)], Takase Shimizu[1,2], Takafumi Akiho[1], Koji Muraki[1], and Norio Kumada[1]

[1]NTT Basic Research Laboratories, NTT Corporation, 3-1 Morinosato-Wakamiya, Atsugi, Kanagawa 243-0198, Japan

[2]Institute for Solid State Physics, The University of Tokyo, 5-1-5 Kashiwanoha, Kashiwa, Chiba 277-8581, Japan

[a)]Current address: VTT Technical Research Centre of Finland, Tietotie 3, 02150 Espoo, Finland

[b)]Author to whom correspondence should be addressed: masayuki.hashisaka.wf@hco.ntt.co.jp

## 1. Homemade HEMT

Supplementary Fig. S1(a) shows an optical micrograph of our HEMT, fabricated in a two-dimensional electron system (electron density of $5.1 \times 10^{11}$ cm$^{-2}$; mobility of $4.1 \times 10^5$ cm$^2$V$^{-1}$s$^{-1}$ at 1.5 K) 55 nm below the surface of a GaAs/AlGaAs heterostructure grown by molecular beam epitaxy (MBE) on a semi-insulating GaAs substrate. We used photolithography for patterning the in-plane semiconductor structure, the Ti/Au Schottky gate, and the ohmic contacts formed by alloying Au-Ge-Ni on the surface. The HEMTs have transimpedance $g_\mathrm{m} \approx 100$ mS at the best operating point in the TIAs. Supplementary Fig. S1(b) presents the input-referred voltage noise $S^\mathrm{V}_\mathrm{HEMT}$ measured for the HEMT in a common-source circuit (inset). The noise property is comparable to the one reported in our previous paper [16].

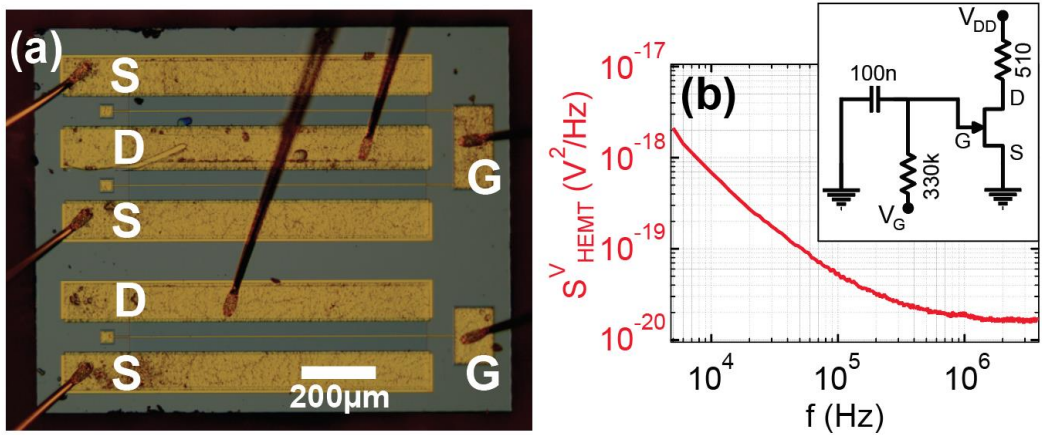

**Supplementary Figure S1**. (a) Optical micrograph of a wire-bonded HEMT chip (H5 in Fig 1 in the main text). We used three HEMTs (standard finger-gate topology) of 1-mm width in parallel so that the total channel width $W = 3$ mm, and gate length $L_\mathrm{g} = 4$ μm. (b) Input voltage-noise spectrum of the HEMT evaluated in the common-source circuit (inset).

## 2. TIA footprint and design

Supplementary Fig. S2 shows an optical micrograph of the TIA constructed on a compact printed circuit board (PCB) packed in the gold-plated copper box (4 × 7 cm$^2$ in size). The TIA is small enough to install several of them either on the 4-K or still plate of a standard commercial dilution refrigerator.

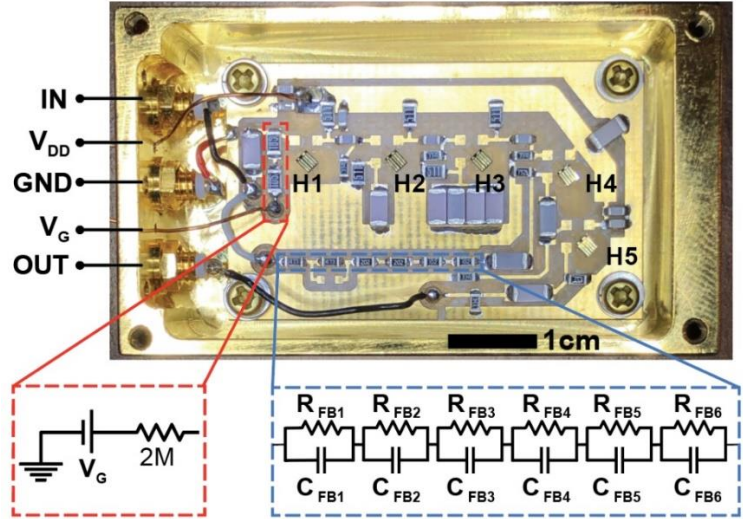

**Supplementary Figure S2**. Optical micrograph of TIA packed into the gold-plated copper box. The copper box has signal input (IN) and output (OUT) ports and a port for the cold ground line (GND). The electronic circuit is assembled on the gold-plated PCB with SMD resistors, capacitors, and homemade HEMTs. Red and blue dashed squares highlight the main differences from our previous TIA, as explained in the main text.

## 3. Feedback resistance

Supplementary Table S1 summarizes the nominal values of SMD resistors $R_{FB1}$–$R_{FB6}$ used in the TIAs with $R_{FB}$ = 50, 100, 200, and 400 kΩ. Although $R_{FB}$ = 200 and 400 kΩ are composed of 206 and 406 kΩ total resistance, we refer to them as 200 and 400 kΩ in the main text for simplicity.

| $R_{FB}$ | 50 kΩ | 100 kΩ | 200 kΩ | 400 kΩ |
|---|---|---|---|---|
| $R_{FB1}$ (kΩ) | 22 | 47 | 100 | 200 |
| $R_{FB2}$ (kΩ) | 22 | 47 | 100 | 200 |
| $R_{FB3}$ (kΩ) | 2 | 2 | 2 | 2 |
| $R_{FB4}$ (kΩ) | 2 | 2 | 2 | 2 |
| $R_{FB5}$ (kΩ) | 1 | 1 | 1 | 1 |
| $R_{FB6}$ (kΩ) | 1 | 1 | 1 | 1 |

**Supplementary Table S1**. $R_{FB1}$-$R_{FB6}$ used in the TIA circuits.

## 4. Stability and reproducibility of TIA characteristics

Our TIA is stable against thermal cycling, time-lapse, and several sweeps for $V_{DD}$ and $V_G$. The stability is confirmed in Supplementary Fig. S3 that shows the total supply current $I_{DD}$ for two both-way sweeps of $V_{DD}$ from 0 V to 1 V at $V_G = -0.31$ V. The perfect overlaps of the two sweeps unambiguously indicate the stability of the TIAs. Note that the $I_{DD}$-$V_{DD}$ plots show changes in the slope at $V_{DD} = 0.4$ V and 0.7 V, reflecting the pinch-off of H1–H3 in the amplification stages and H4 and H5 in the source-follower stages.

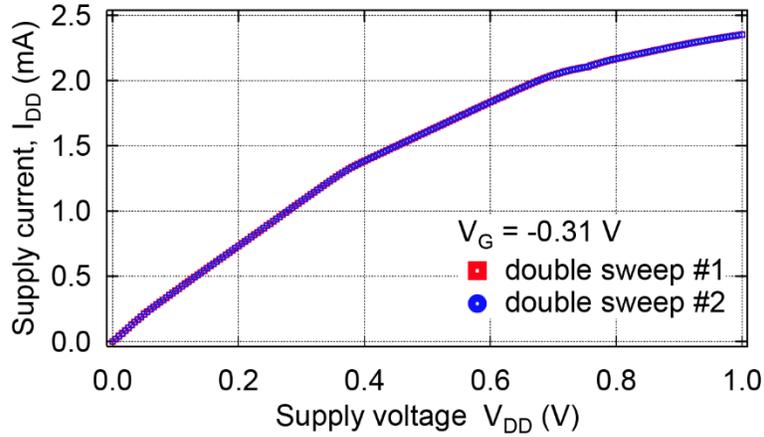

**Supplementary Figure S3.** Total supply current $I_{DD}$ for two both-way sweeps of $V_{DD}$ from 0 V to 1 V at $V_G, = -0.31$ V. (TIA with $R_{FB} = 100$ kΩ). Power consumption at the typical operation point $V_{DD} = 0.7$ V is given by 0.7 V × 2.1 mA ≈ 1.5 mW.

## 5. Operating TIA with finite input capacitance

Generally, $C_{IN}$ induces a phase shift in the feedback loop to destabilize the TIA operation and increase $S_{input}$ at high frequencies. To examine the impact of $C_{IN}$, we examined operating the TIA with $C_{IN} = 30$ pF (comparable to the parasitic capacitance of approximately 50-cm-long coaxial cable) and compared the result with $C_{IN} \approx 0$ pF, which is the case discussed in the main text. Supplementary Fig. S4 presents the frequency $f$ dependence of $|Z_{trans}|$ (upper) and $S_{input}$ (lower) of the $R_{FB} = 200$ kΩ TIAs with $C_{IN} \approx 0$ pF (pink) and 30 pF (black). While both TIAs show similar characteristics at $f < 1$ MHz, $|Z_{trans}|$ decreases and $S_{input}$ increases more rapidly with increasing $f$ for $C_{IN} = 30$ pF than $C_{IN} \approx 0$ pF, indicating that a finite $C_{IN}$ degrades the TIA's performance (for the detailed mechanism, see Ref. [19]). However, the degradation is tiny, so we consider it is not critical in most experiments. To conclude, practically, $C_{IN}$ of the parasitic capacitance of a coaxial cable connecting a sample and our TIA does not cause serious problems.

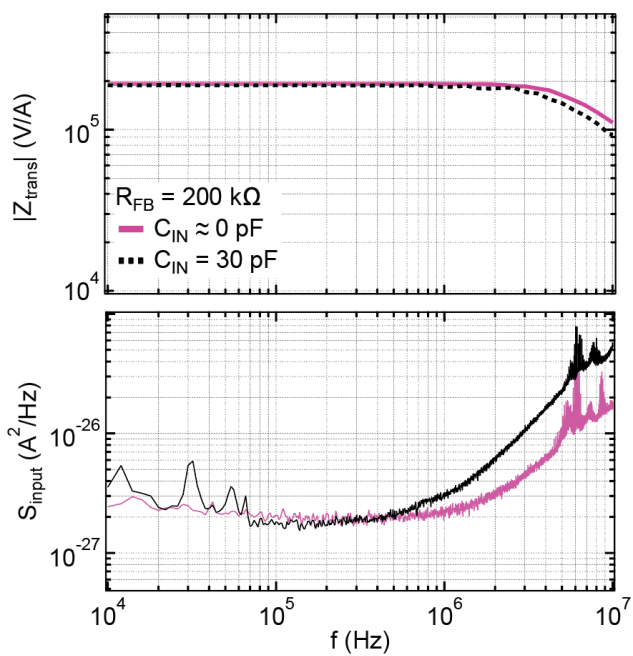

**Supplementary Figure S4.** Comparison of $|Z_{trans}|$ (upper) and $S_{input}$ (lower) between $R_{FB}$ = 200 kΩ TIAs with $C_{IN}$ = 30 pF (black) and $C_{IN} \approx 0$ pF (pink). $C_{IN}$ = 30 pF is comparable to the parasitic capacitance of a coaxial cable connecting a sample at the mixing-chamber stage and a TIA at the still or 4-K stage in a dilution refrigerator.

## 6. Summary of SNR and $\sigma_{av}$ of time-domain current-noise measurements with LPFs

Supplementary Fig. S5 summarizes the SNR and $\sigma_{av}$ of the 2-nA peak-to-peak square-wave measurements performed with different LPFs (for 1 MHz, Stanford Research Systems SR560; from 1.9 MHz to 45 MHz, Mini-Circuits BLP Series).

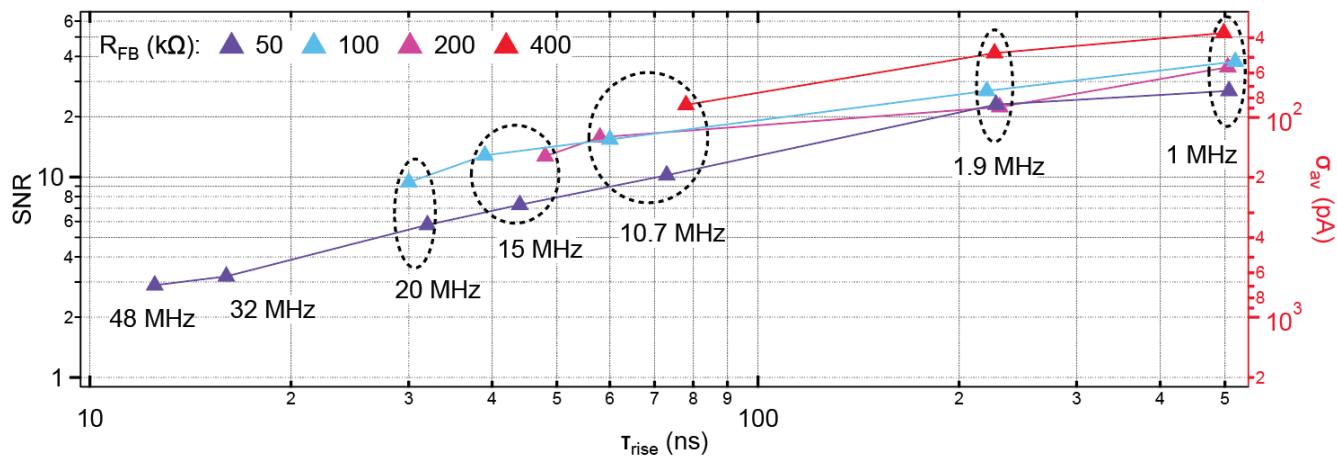

**Supplementary Figure S5.** The SNR (left) and $\sigma_{av}$ (right) values of the TIAs obtained with different LPF frequencies. The dashed black lines are used to group the data points obtained at the same LPF frequency.